\documentclass[aps,pra,onecolumn,citeautoscript,superscriptaddress,nofootinbib,preprintnumbers]{revtex4-1}  
\usepackage{amsmath,amssymb,bm} 
\usepackage{graphicx}  
\usepackage{color}
\usepackage[papersize={8.5in,11in}]{geometry}
\usepackage[colorlinks=true]{hyperref}  
\hypersetup{
    unicode=false,          
    pdftoolbar=true,        
    pdfmenubar=true,        
    pdffitwindow=false,     
    pdfstartview={FitH},    
    pdftitle={A critical strange metal from fluctuating gauge fields in a solvable random model},    
    pdfauthor={Aavishkar A. Patel},     
    pdfsubject={},   
    pdfcreator={},   
    pdfproducer={}, 
    pdfkeywords={} {} {}, 
    pdfnewwindow=true,      
    colorlinks=true,       
    linkcolor=magenta, 
    citecolor=blue,        
    filecolor=magenta,      
    urlcolor=blue           
} 

\usepackage{amsfonts}
\usepackage{upgreek}
\usepackage{slashed}
\usepackage{latexsym}

\newcommand{\beq}{\begin{equation}}
\newcommand{\eeq}{\end{equation}}
\newcommand{\nn}{\nonumber \\}

\def \dg{\dagger}
\def \pr{\prime}

\date{\today}

\begin{document}

\title{A critical strange metal from fluctuating gauge fields in a solvable random model}

\author{Aavishkar A. Patel}
\affiliation{Department of Physics, Harvard University, Cambridge MA 02138, USA}

\author{Subir Sachdev}
\affiliation{Department of Physics, Harvard University, Cambridge MA 02138, USA}
\affiliation{Perimeter Institute for Theoretical Physics, Waterloo, Ontario, Canada N2L 2Y5}

\begin{abstract}
Building upon techniques employed in the construction of the Sachdev-Ye-Kitaev (SYK) model, which is a solvable $0+1$ dimensional model of a non-Fermi liquid, we develop a solvable, infinite-ranged random-hopping model of fermions coupled to fluctuating U(1) gauge fields. In a specific large-$N$ limit, our model realizes a gapless non-Fermi liquid phase, which combines the effects of hopping and interaction terms. We derive the thermodynamic properties of the non-Fermi liquid phase realized by this model, and the charge transport properties of an infinite-dimensional version with spatial structure. 
\end{abstract}

\maketitle

\section{Introduction}
\label{intro}

A number of models of strange metals have been been constructed \cite{PG98,Gu17,Gu2017local,Sachdev2017,Balents2017,Zhang17,Shenoy2018,McGreevy2017,Patel2018,Chowdhury2018,FGST18} by connecting together `quantum islands', in which each island has random all-to-all interactions between the electrons {\em i.e.\/} each island is a 0+1 dimensional SYK model \cite{Sachdev1993,kitaev2015talk,Sachdev2015,KitaevSuh}. Some of these models \cite{PG98,Balents2017,Patel2018,Chowdhury2018} exhibit `bad metal' behavior above some crossover temperature, with a resistivity which increases linearly with temperature ($T$), and has a magnitude (in two dimensions) which is larger than the quantum unit of resistance $h/e^2$. These models can be useful starting points for understanding a variety of experiments above
moderate values of $T$, and they also predict \cite{Sachdev1993,PG98} the frequency independent 
density fluctuation spectrum observed in recent electron scattering experiments \cite{Abbamonte17}.
However, some of the most interesting and puzzling observations exhibit \cite{Hussey09,Jin11,Legros18} linear-in-$T$ resistivity down to vanishingly
small $T$, with a resistivity which is much smaller than $h/e^2$. Kondo-like two-band SYK models have been proposed for such behavior \cite{Patel2018,Chowdhury2018}, in which a band of itinerant electrons acquires marginal-Fermi liquid behavior \cite{Varma89} upon Kondo exchange scattering off localized electrons in SYK islands. 
The holographic models of strange metals have a structure very similar to these Kondo-SYK models \cite{Faulkner2013,Cubrovic:2009ye,SS10}.

A possible shortcoming of the two-band SYK-Kondo models \cite{Patel2018,Chowdhury2018} is that density of itinerant carriers is `small'. In other words, only the itinerant electrons 
carry current and exhibit marginal-Fermi liquid behavior, while the localized electrons in SYK islands only act as a background `bath' of incoherent electrons which dissipates current from the itinerant electrons. This behavior does not appear to be in accord with estimates of the magnitude of the linear-in-$T$ resistivity as $T \rightarrow 0$ \cite{Legros18}.

\begin{figure}[h]
\begin{center}
\includegraphics[height=2.5in]{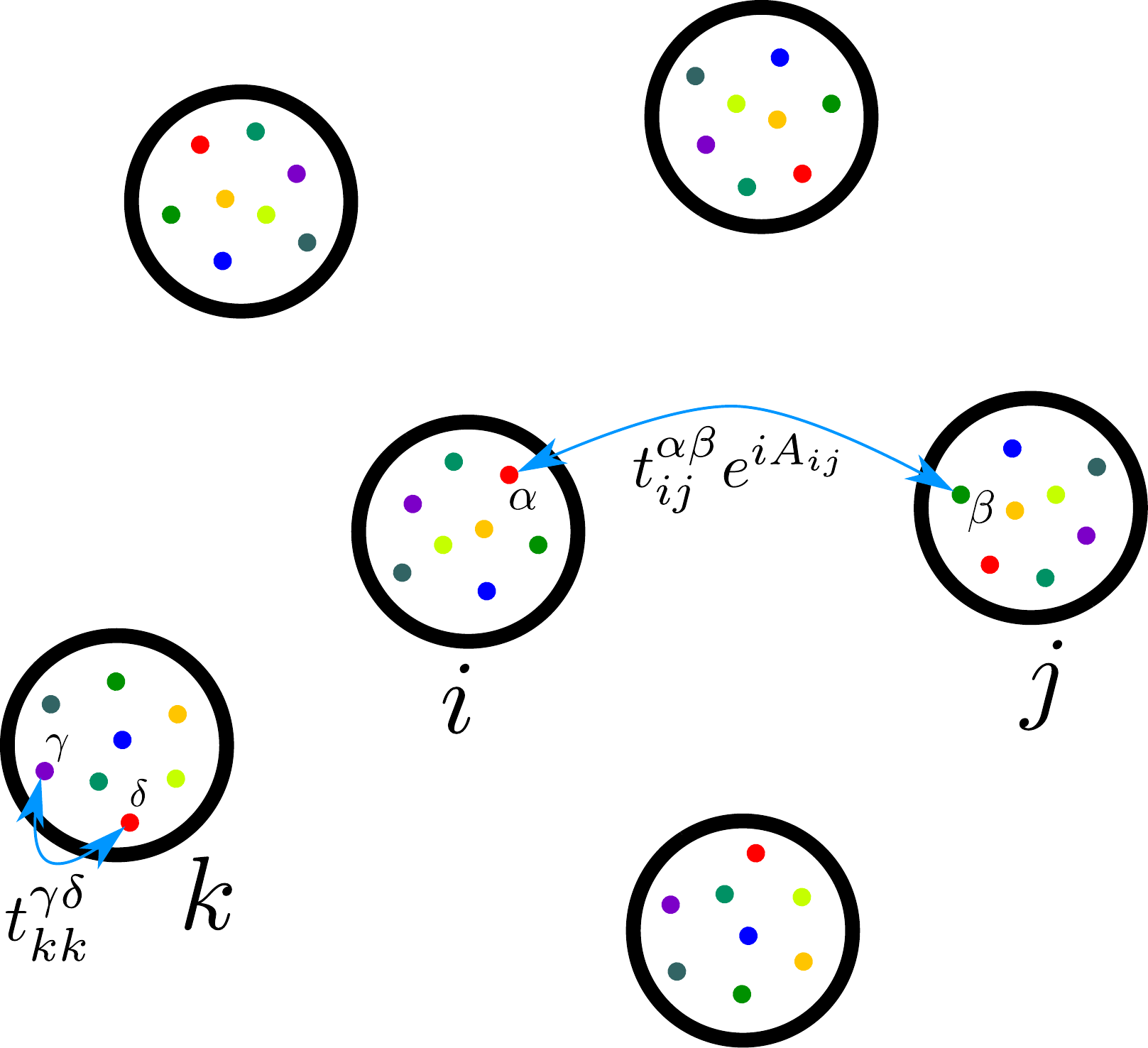} 
\end{center}
\caption{A cartoon of our model. It consists of $N$ clusters indexed by $i,j,...$, each of which contains $M$ sites indexed by $\alpha,\beta,...$. Random hopping occurs between all possible pairs of intra-cluster and inter-cluster sites, but only inter-cluster hops are coupled to dynamic U(1) gauge fields $A_{ij}$. The model is solved in the $M,N\rightarrow\infty$ limit, with $M/N$ fixed.}
\label{cartoon}
\end{figure}

In this paper, we shall propose and solve a SYK-like model which exhibits strange metal resistivity as $T \rightarrow 0$, and in which the density of itinerant fermions is `large'. 
We shall examine a model of fermions coupled to an emergent, dynamic, U(1) gauge field.
We shall show that a solvable SYK-like large $N$ limit exists, in which the electrons are in $N$ clusters with $M$ sites per cluster ($M/N$ is fixed as the large $N$ limit is taken): see Fig.~\ref{cartoon}.
The DC conductivity of our model is presented in Eq.~(\ref{sigDC}), and the resistivity varies as $T^{2x}$ as $T \rightarrow 0$, with the exponent $x$ dependent only upon $M/N$ and the particle-hole asymmetry parameter $\mathcal{E}$, as shown in Eq.~(\ref{xasym}) and Fig.~\ref{exponent}. In the limit of small $M/N$, $2x \sim 1$ (see Fig.~\ref{exponent}), 
and then we have nearly linear-in-$T$ resistivity.

The problem of a finite density of fermions coupled to an emergent gauge field appears in many different physical contexts. The most extensively studied case is that related to compressible quantum Hall states in a half-filled Landau level \cite{HLR}. These studies begin with assumption that the fermions form a Fermi surface, and Landau damping from the fermions leads to an overdamped gauge propagator. The effects of the gauge coupling and the disorder are then treated perturbatively. The presence of disorder has a relatively modest effect in inducing a diffusive form for the gauge propagator. In the present paper we shall take a random all-to-all form of the fermion propagator, and show that this allows for an exact treatment of the gauge fluctuations.
The local criticality exhibited by our model is expected to eventually crossover at low enough $T$ to more generic finite-dimensional behavior, but there is no theory yet for such a fixed point with strong disorder and interactions.

The physical context most appropriate for our proposed connection to observations on the overdoped cuprates \cite{Hussey09,Legros18} is the theory of an `algebraic charge liquid' (ACL) \cite{Kaul2008} of spinless fermionic chargons coupled to an emergent gauge field. Specifically, in a SU(2) gauge theory of optimal doping quantum criticality \cite{SS09,Chowdhury2015,SSNambu,Scheurer18}, it has been proposed that there could be an overdoped phase with a large density of fermionic chargons coupled to a deconfined SU(2) gauge field. For simplicity, this paper will consider the U(1) gauge field case, although the properties of the SU(2) case are expected to be very similar. 

We will begin in Section~\ref{model} by defining the model and computing its saddle point equations in the large $N$ limit. The properties of the single fermion Green's function as a function of frequency, temperature, and chemical potential will be described in Section~\ref{sec:single}.
The thermodynamics will be described in Section~\ref{thermo}, and we will describe a higher-dimensional generalization which allows us to compute transport properties in Section~\ref{transport}.

Appendix~\ref{higgs} describes an extension of our model in which the condensation of a charge 2 Higgs field leads to a metallic phase in which the fermions carry $\mathbb{Z}_2$ gauge charges. The Higgs condensate quenches the gauge field fluctuations, and the transport is therefore Fermi-liquid like.
The Higgs condensate also reduces the density of low-energy fermionic excitations, and so we may view this transition as a model \cite{SS09,Chowdhury2015,SSNambu,Scheurer18} of optimal doping criticality from the overdoped side (no Higgs condensate) to the underdoped side (Higgs condensate present).

\section{Model and large-$N$ limit}
\label{model}

We study a model of $N$ clusters, each with $M$ flavors of fermions, with infinite-ranged random hopping between the clusters that is coupled to fluctuating U(1) gauge fields. It is given by  
\beq
\mathcal{H} = -\frac{1}{(MN)^{1/2}}\sum_{ij=1}^N\sum_{\alpha\beta=1}^M \left[t_{ij}^{\alpha \beta}e^{iA_{ij}} f^\dg_{i\alpha}f_{j\beta}+(MN)^{1/2}\mu \delta_{ij}^{\alpha\beta} f^\dg_{i\alpha}f_{i\alpha}\right],~~\ll t_{ij}^{\alpha \beta} t_{ji}^{\beta \alpha} \gg =\ll |t_{ij}^{\alpha \beta}|^2\gg = t^2,~~A_{ji}=-A_{ij}.
\label{H}
\eeq
where $N,M\rightarrow \infty$ and $M/N$ is an $\mathcal{O}(1)$ quantity. The $t_{ij}^{\alpha\beta}$ are complex gaussian random variables and $\ll..\gg$ denotes disorder-averaging; all disorder averages other than the ones explicitly shown above are zero. The clusters are indexed by $i$, and the sites (flavors) within a cluster, are indexed by $\alpha$. A cartoon of our model is shown in Fig.~\ref{cartoon}. 

As in the analysis of the SYK models~\cite{Sachdev2015,Sachdev2017}, we average over realizations of disorder. Doing so formally requires introducing replicas; however we assume, like in the SYK models, that there is no replica-symmetry breaking, restricting to replica-diagonal configurations and suppressing the then trivial sum over replicas. We introduce bilocal (in time) fields $G$ and $\Sigma$, obtaining the Euclidean action 
\begin{align}
&S = \int d\tau \sum_{i=1}^N\sum_{\alpha=1}^M f^\dg_{i\alpha}(\tau)(\partial_\tau+iA^0_i(\tau)+\mu)f_{i\alpha}(\tau) +t^2\frac{M}{N}\int d\tau d\tau^\pr \sum_{ij=1,i\le j}^N e^{i(A_{ij}(\tau)-A_{ij}(\tau^\pr))}G_j(\tau-\tau^\pr)G_i(\tau^\pr-\tau) \nn
&-M\int d\tau d\tau^\pr \sum_{i=1}^N\Sigma_i(\tau-\tau^\pr)\left[G_i(\tau^\pr-\tau)-\frac{1}{M}\sum_{\alpha=1}^Mf_{i\alpha}(\tau^\pr)f^\dg_{i\alpha}(\tau)\right].
\label{S1}
\end{align}
The partition function is given by $Z=\int \mathcal{D}f\mathcal{D}f^\dg\mathcal{D}A\mathcal{D}G\mathcal{D}\Sigma~e^{-S}$, and $\tau$ denotes Euclidean time. Unbounded integrals denote integration over the full range of the pertinent variable. Integrating out the Lagrange multipliers $\Sigma_i$ followed by the $G_i$ restores the pure disorder-averaged action. In the $M\rightarrow\infty$ limit, the integrals over the $\Sigma_i$ enforce the definitions of $G_i$ on each cluster $i$. The disorder averaged action is gauge-invariant under the transformations 
\beq
A_{ij}(\tau)\rightarrow A_{ij}(\tau)+\theta_i(\tau)-\theta_j(\tau),~~f_{i\alpha}(\tau)\rightarrow f_{i\alpha}(\tau)e^{i\theta_i(\tau)},~~A^0_i(\tau)\rightarrow A^0_i(\tau)-\partial_\tau\theta_i(\tau),
\label{GT}
\eeq
with $G_i(\tau-\tau^\pr)\rightarrow G_i(\tau-\tau^\pr)e^{i(\theta_i(\tau)-\theta_i(\tau^\pr))}$ and $\Sigma_i(\tau-\tau^\pr)\rightarrow \Sigma_i(\tau-\tau^\pr)e^{i(\theta_i(\tau)-\theta_i(\tau^\pr))}$. The propagators of the scalar potentials $A^0_i$ will be screened due to the finite density of fermions~\cite{Kim1994}; fluctuations of the $A^0_i$ will be hence unable to inflict any singular self energy on the fermions at low energies, and we will thus simply ignore the $A^0_i$.  

Examining the disorder-averaged action, after integrating out the fermions, does not immediately suggest a large-$N$ saddle-point for the $G_i$, but a simple large $N$ limit does turn out to exist. The reason is that there are enough ($M$) sites per cluster to self-average the cluster Green's function $G_i$, so that the solution will have $G_i$ that don't depend on $i$, even though there are $N$ clusters. This can be seen easily when the coupling to the gauge fields is turned off. Then we know the standard result for the fully-averaged Green's function $G_{\mathrm{avg}}$ of the full large-$MN$ random matrix {\it exactly}, but can also express it as
\beq
G_{\mathrm{avg}}(\tau-\tau^\pr) = \frac{1}{MN}\sum_{i=1}^N\sum_{\alpha=1}^M\langle f_{i\alpha}(\tau)f^\dg_{i\alpha}(\tau^\pr)\rangle = \frac{1}{N} \sum_i G_i(\tau-\tau^\pr). 
\eeq
Then, the second term of (\ref{S1}) may be written as
\beq
M\frac{t^2}{2}\int d\tau d\tau^\pr G_{\mathrm{avg}}(\tau-\tau^\pr) \sum_{i=1}^N G_i(\tau^\pr-\tau).
\eeq
Since there are now appropriate prefactors of $M$ everywhere in all terms in $S$ after integrating out the fermions, we can take functional derivatives with respect to $G_i$ and $\Sigma_i$  (remembering that $G_{\mathrm{avg}}$ contains $G_i$) and write down the saddle-point $\Sigma_i(\tau-\tau^\pr) = t^2 G_{\mathrm{avg}}(\tau-\tau^\pr)$ and $G_i(i\omega_n)=1/(i\omega_n+\mu-\Sigma_i(i\omega_n))$, which are independent of $i$, indicating that the cluster-averaged (over $M$ sites) Green's function is the same as the fully averaged (over $MN$ sites and clusters) Green's function at large-$M,N$. Another way to see this qualitatively is that the distribution for $G$'s averaged over $M$ sites is the convolution of $M$ distributions for the single-site $G$'s. For Gaussians, this would imply that its variance is $1/M^{\mathrm{th}}$ of that of the single-site distribution, which, although much larger than the variance of the fully averaged $G$ (which is $1/(MN)^{\mathrm{th}}$ of that of the single-site distribution), should still be small as $M\rightarrow\infty$. 

Turning the gauge fields back on, we expand out the exponentials to quadratic order (assuming that monopoles are irrelevant and there is no confinement transition, so the compactness of the gauge fields isn't important; we will discuss this further at the end of Sec.~\ref{thermo}) and obtain, 
\begin{align}
&S = \int d\tau \sum_{i=1}^N\sum_{\alpha=1}^M f^\dg_{i\alpha}(\tau)(\partial_\tau+iA^0_i(\tau)+\mu)f_{i\alpha}(\tau) \nn
&+t^2\frac{M}{N}\int d\tau d\tau^\pr \sum_{ij=1,i\le j}^N \left[1+i(A_{ij}(\tau)-A_{ij}(\tau^\pr))-\frac{1}{2}A_{ij}^2(\tau)-\frac{1}{2}A_{ij}^2(\tau^\pr)+A_{ij}(\tau)A_{ij}(\tau^\pr)\right]G_j(\tau-\tau^\pr)G_i(\tau^\pr-\tau) \nn
&-M\int d\tau d\tau^\pr \sum_{i=1}^N\Sigma_i(\tau-\tau^\pr)\left[G_i(\tau^\pr-\tau)-\frac{1}{M}\sum_{\alpha=1}^Mf_{i\alpha}(\tau^\pr)f^\dg_{i\alpha}(\tau)\right].
\end{align}
This expanded-out action is also gauge-invariant under the previously mentioned transformation, up to quadratic order in the gauge fields and their shifts. The terms linear in $A_{ij}$ in the second line of the above vanish, and the $A^2_{ij}$ terms can be reorganized, 
\begin{align}
&S = \int d\tau \sum_{i=1}^N\sum_{\alpha=1}^M f^\dg_{i\alpha}(\tau)(\partial_\tau+iA^0_i(\tau)+\mu)f_{i\alpha}(\tau) + \frac{T}{2}\sum_{\Omega_m}\sum_{ij=1,i\le j}^N A_{ij}(i\Omega_m)\left[\Pi_{ij}(i\Omega_m)-\Pi_{ij}(i\Omega_m=0)\right]A_{ij}(-i\Omega_m) \nn
&+t^2\frac{M}{N}\int d\tau d\tau^\pr \sum_{ij=1,i\le j}^N G_j(\tau-\tau^\pr)G_i(\tau^\pr-\tau) \nn
&-M\int d\tau d\tau^\pr \sum_{i=1}^N\Sigma_i(\tau-\tau^\pr)\left[G_i(\tau^\pr-\tau)-\frac{1}{M}\sum_{\alpha=1}^Mf_{i\alpha}(\tau^\pr)f^\dg_{i\alpha}(\tau)\right],
\label{S2}
\end{align}
with 
\beq
\Pi_{ij}(i\Omega_m)=2t^2\frac{M}{N}\int d\tau e^{i\Omega_m\tau} G_i(\tau)G_j(-\tau).
\eeq 

\begin{figure}
\begin{center}
\includegraphics[height=1.8in]{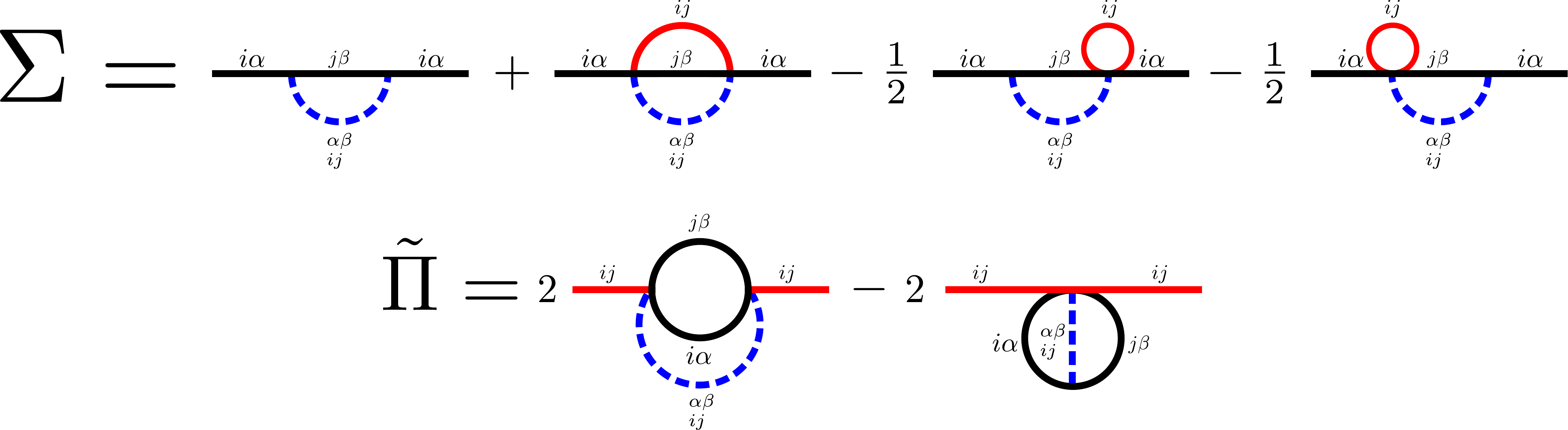} 
\end{center}
\caption{Diagrammatic representation of the fermion ($\Sigma$) and regularized gauge field ($\tilde{\Pi}=\Pi(i\Omega_m)-\Pi(i\Omega_m=0)$) self-energies for the Dyson equation (\ref{Dyson1}). The black lines are fermion propagators, the red lines are gauge field propagators, and the dashed blue lines are contractions of the gaussian random variables $t^{\alpha\beta}_{ij}$ coming from the disorder average. These are the only diagrams that contribute in in the large-$M,N$ limit.}
\label{gfdiagrams}
\end{figure}

We proceed to integrate out the fermions and the gauge fields. Normally, integrating out the gauge fields requires gauge-fixing in order to avoid overcounting redundant configurations. However, in the large-$N$ limit here, we have $\mathcal{O}(N^2)$ gauge variables $A_{ij}$, but only $\mathcal{O}(N)$ constraining variables $\theta_i$. The space of gauge field configurations is then $\sim\mathbb{R}^{N^2}$, whereas the space occupied by configurations redundant to a single configuration, generated by shifting the $\mathcal{O}(N^2)$ $A_{ij}$'s by $N$ $\theta_i$'s is $\sim\mathbb{R}^N$. Therefore the space of unique gauge configurations is $\sim\mathbb{R}^{N^2}/\mathbb{R}^N$, which at leading order in large-$N$ is approximately $\mathbb{R}^{N^2}$. Thus, we can just naively integrate out the $A_{ij}$ in the large-$N$ limit, and the corrections from gauge-fixing will not affect the free energy and the saddle-point values of $G$ and $\Sigma$ at leading order in the large-$N$ limit. After integrating out, we obtain
\begin{align}
&TS = -MT\sum_{\omega_n}\sum_{i=1}^N\ln\left[i\omega_n+\mu-\Sigma_i(i\omega_n)\right]+ \frac{T}{2}\sum_{\Omega_m\neq0}\sum_{ij=1,i<j}^N \ln\left[\Pi_{ij}(i\Omega_m)-\Pi_{ij}(i\Omega_m=0)\right] \nn
&+t^2\frac{M}{N} T\sum_{\omega_n}\sum_{ij=1,i\le j}^N G_j(i\omega_n)G_i(i\omega_n) -M T\sum_{\omega_n} \sum_{i=1}^N\Sigma_i(i\omega_n)G_i(i\omega_n).
\label{S2intout}
\end{align}
where, as mentioned earlier, we neglect the time components of the gauge fields. Varying with respect to $G_i(i\omega_n)$ and $\Sigma_i(i\omega_n)$, produces a site-uniform saddle-point described by (after dropping site-dependent subscripts)
\begin{align}
&\Sigma(i\omega_n)=t^2G(i\omega_n) +t^2T\sum_{\Omega_m\neq0}\frac{G(i\omega_n+i\Omega_m)-G(i\omega_n)}{\Pi(i\Omega_m)-\Pi(i\Omega_m=0)}, \nn
&\Pi(i\Omega_m) = 2t^2T\frac{M}{N}\sum_{\omega_n}G(i\omega_n)G(i\omega_n+i\Omega_m),~~G(i\omega_n)  = \frac{1}{i\omega_n+\mu-\Sigma(i\omega_n)}.
\label{Dyson1}
\end{align}
These equations can also be derived diagrammatically starting from (\ref{H}) in the large-$M,N$ limit, and expanding the exponential to quadratic order after disorder-averaging (Fig.~\ref{gfdiagrams}).

Note that the zero Matsubara frequency component of $A_{ij}$ does not contribute to the action (\ref{S2}) or (\ref{S2intout}) even at $T\neq0$. The gauge field contribution to the fermion self energy $\Sigma(i\omega_n)$ in (\ref{Dyson1}) thus doesn't involve the zero Matsubara frequency component of the gauge field propagator. This is because, as far as the fermions are concerned, the zero Matsubara frequency components are just static phase shifts of the $t_{ij}^{\alpha\beta}$, and have already been accounted for while disorder averaging. This absence of the zero frequency components has consequences for the thermodynamic properties of the saddle-point solution, and certain modifications have to be made to ensure that the saddle-point is thermodynamically stable (see Sec.~\ref{thermo}). However, these modifications do not affect the saddle-point solution to be detailed in the next section above some energy scale which can be made arbitrarily small. 

If we consider fluctuations $(\delta G_i(i\omega_n),\delta \Sigma_i(i\omega_n))$ about the saddle-point action that do not amount to simply changing a gauge, the kernel of their action at quadratic order is given by $\hat{K}_{ij} = \hat{K}^{(1)}\delta_{ij}+\hat{K}^{(2)}$, where $\hat{K}^{(1,2)}$ are matrices in $(\delta G,\delta \Sigma)$ and frequency space. Here $\hat{K}^{(1)}$ is of order $M$, coming from the fermion determinant and $\Sigma G$ terms of (\ref{S2intout}), and $\hat{K}^{(2)}$, which comes from the other two terms is of order $1$. Then, diagonalizing $\hat{K}$ in $i,j$ and $(\delta G,\delta \Sigma)$ space produces $\mathcal{O}(N)$ fluctuation eigenmodes with eigenvalues that are $\mathcal{O}(M)$. Integrating over these $N$ modes yields a sub-leading $\mathcal{O}(N)$ contribution to the free energy, and each of these modes also has an $\mathcal{O}(M)$ stiffness that suppresses its fluctuations. Hence, the saddle-point described by (\ref{Dyson1}) is well-defined.

\section{Single-particle properties}
\label{sec:single}

\subsection{Zero temperature}
\label{gauge0}

\begin{figure}
\begin{center}
\includegraphics[height=2.0in]{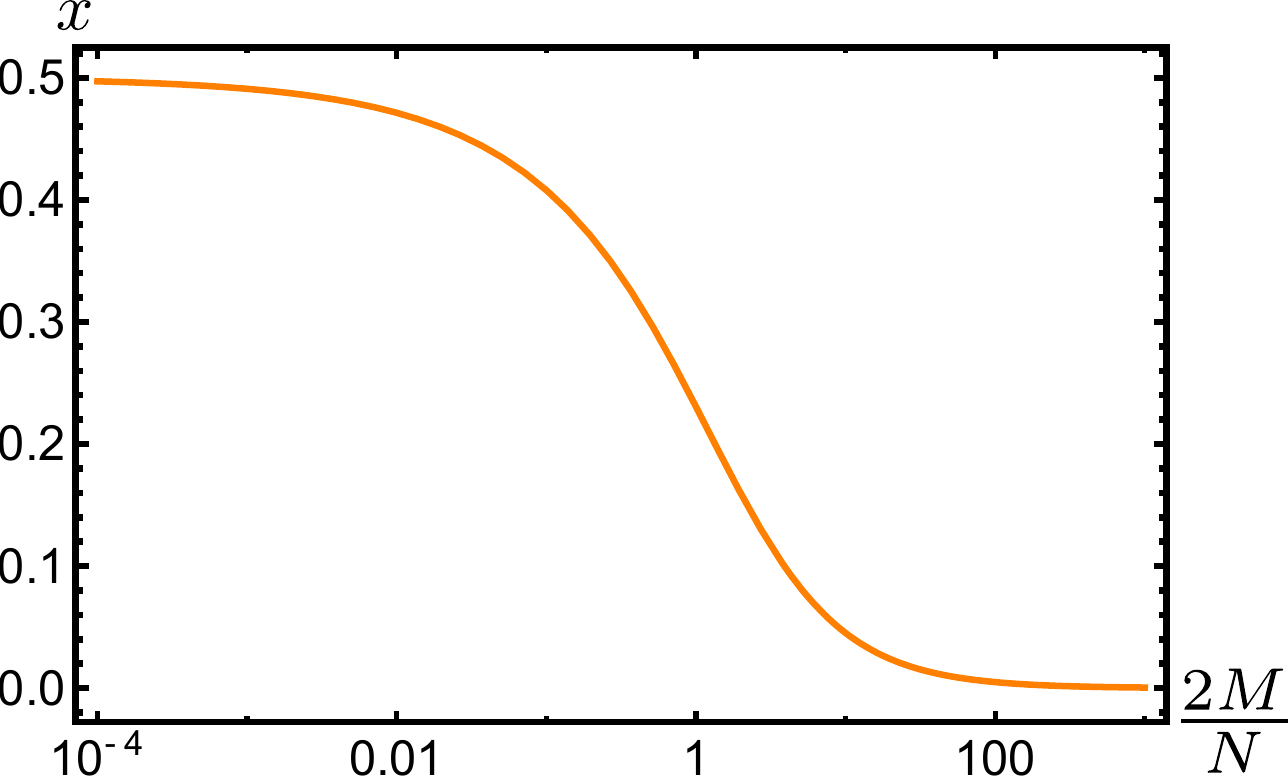} 
\end{center}
\caption{Plot of the exponent $x$ giving the frequency scaling of the IR fermion self-energy, vs $2M/N$, at half-filling.}
\label{exponent}
\end{figure}

We solve for the fermion and gauge field propagators at $T=0$. We set $\mu=0$ (corresponding to half-filling, see Sec.~\ref{notahalf} for $\mu\neq0$), and start with an ansatz for $G$ in the IR at $T=0$,
\beq
G(\tau) = -C\frac{\mathrm{sgn}(\tau)}{t^{1-x}|\tau|^{1-x}},~~G(i\omega_n) = -2 i C t^{x-1}\sin \left(\frac{\pi  x}{2}\right) \Gamma (x) \mathrm{sgn}(\omega_n) |\omega_n|^{-x},~~ 0<x<\frac{1}{2},~~C>0.
\label{gIR}
\eeq
We then obtain 
\beq
\Pi(i\Omega_m)-\Pi(i\Omega_m=0) = -4(M/N)C^2 t^{2x}\sin (\pi  x) \Gamma (2 x-1) \left|\Omega_m\right|^{1-2x}.
\label{pIR}
\eeq
This the the fermion self-energy
\begin{align}
&\Sigma(i\omega_n) = \frac{iN\sqrt{\pi } 2^{2 x-1} \sin \left(\frac{\pi  x}{2}\right) \csc ^2(\pi x)}{2M C x \Gamma (2 x-1) \Gamma \left(\frac{1}{2}-x\right)}\mathrm{sgn}(\omega_n) t^{1-x}\left|\omega_n\right|^{x} + t^2G(i\omega_n)\left[1-\int\frac{d\Omega_m}{2\pi}\frac{1}{\Pi(i\Omega_m)-\Pi(i\Omega_m=0)}\right].
\label{fsigma}
\end{align}
The integral over $\Omega_m$ contains contributions from frequencies outside the regime of validity of the IR solution, and hence requires a UV completion in order to be evaluated. We assume that the UV completion is such that the term in square brackets evaluates to zero, which we will justify below; the vanishing of the square bracketed term is also confirmed by our numerical analysis of the UV complete theory below. Then, using $G(i\omega_n)=-1/\Sigma(i\omega_n)$, we find that we cannot determine $C$ (it cancels between the LHS and RHS of the equation), but we can determine the universal exponent $x$ by solving
\beq
\frac{1/x-2}{1+\sec (\pi  x)} = \frac{2M}{N},
\label{expeq}
\eeq
with $x$ vs $2M/N$ plotted in Fig.~\ref{exponent}. The fact that we can't determine $C$ purely from the IR properties indicates that it is non-universal.
 
We now justify the vanishing of the term in square brackets in (\ref{fsigma}): Suppose it didn't exactly vanish, and $\int d\Omega_m/(\Pi(i\Omega_m)-\Pi(i\Omega_m=0))=1-\nu$, where $\nu\ll 1$. Then, this leaves behind a term $\nu t^2G(i\omega_n)$ in the expression for $\Sigma(i\omega_n)$, which, scaling as $\mathrm{sgn}(\omega_n)|\omega_n|^{-x}$, is more relevant at low energies than the other term in $\Sigma(i\omega_n)$. We can then try to ignore the other term in the IR. The Dyson equation becomes 
\beq
\Sigma(i\omega_n) = \nu t^2G(i\omega_n),~~G(i\omega_n) = \frac{1}{i\omega_n-\Sigma(i\omega_n)}, 
\eeq
This equation is solved in the IR by the random-matrix solution $G(i\omega_n) = -i\mathrm{sgn}(\omega_n)/(\nu^{1/2}t)$. This solution then modifies the gauge field propagator in the IR, with $\Pi_{\mathrm{new}}(i\Omega_m)-\Pi_{\mathrm{new}}(i\Omega_m=0) = (2/\pi)(M/N)(|\Omega_m|/\nu)$.
We can then write using (\ref{Dyson1})
\beq
\Sigma(i\omega_n)=-i(t/\nu^{1/2})\mathrm{sgn}(\omega_n)+ i \nu^{1/2} t\int_{-\Lambda}^{\Lambda}\frac{d\Omega_m}{2\pi}\frac{\mathrm{sgn}(\omega_n)-\mathrm{sgn}(\omega_n+\Omega_m)}{(2/\pi)(M/N)|\Omega_m|},
\eeq
where $\Lambda$ is a ``critical window" over which the IR solution is valid. This then gives a singular self-energy 
\beq
\Sigma(i\omega_n)=-i(t/\nu^{1/2})\mathrm{sgn}(\omega_n)\left(1+\frac{N\nu}{2M}\ln\left(\frac{|\omega_n|}{\Lambda}\right)\right) \rightarrow  -i(t/\nu^{1/2})\mathrm{sgn}(\omega_n)\left(\frac{|\omega_n|}{\Lambda}\right)^{\frac{N\nu}{2M}}, 
\label{sigrn}
\eeq
We have thus recovered a power-law self-energy without explicitly assuming $\nu=0$ to begin with. Repeated iterations of (\ref{Dyson1}) then converge the exponent of the power-law to the value defined by (\ref{expeq}).

The Dyson equations (\ref{Dyson1}) are not fully UV-complete, and do not contain enough information to determine the gauge field propagator at high frequencies. In order to solve them numerically, we need a UV-complete set of equations. We do this by adding a ``Maxwell" term to the gauge field action
\beq
S \rightarrow S + \frac{1}{2g^2}\int d\tau \sum_{ij=1,i\le j}^N (\partial_\tau A_{ij}(\tau)+A^0_i(\tau)-A^0_j(\tau))^2,
\label{minUVc}
\eeq
with a gauge coupling $g$, and the $A^0_i$'s may be ignored due to the aforementioned screening. This then adds a term $\Omega_m^2/g^2$ to $\Pi(i\Omega_m)-\Pi(i\Omega_m=0)$ in (\ref{Dyson1}). Note that (\ref{minUVc}) contains only ``electric" kinetic terms for the gauge fields and no ``magnetic" terms that are functions of the sums of gauge link variables $A_{ij}$ around closed loops.  We will discuss the effects of adding magnetic terms in Sec.~\ref{thermo}.
  
The numerical solution was then performed by starting with free fermion and gauge field Green's functions
\beq
G_0(i\omega_n)=\frac{1}{i\omega_n+\mu},~~D_0(i\Omega_m)=\frac{g^2}{\Omega_m^2},
\eeq
and then iterating the Dyson equations (\ref{Dyson1})  in the \verb|MATLAB| code \verb|gd.m|~\cite{gd}. We found that the $t^2G(i\omega_n)$ term in $\Sigma(i\omega_n)$ indeed cancels out as $T\rightarrow 0$, and a power-law scaling of $G(i\omega_n)$ is obtained in the IR, with the exponent given by (\ref{expeq}).  This cancellation of the $t^2G(i\omega_n)$ term holds even for $\mu\neq0$, leading to the results in Sec.~\ref{notahalf}. A $T=0$ numerical solution of the real time version of the Dyson equations (performed in \verb|gdrealtime0.m|~\cite{gdrt}) also yields the appropriate analytically continued version of (\ref{gIR}) for the retarded Green's function in the IR,
\beq
G^R(\omega) = -2 i C t^{x-1}\sin \left(\frac{\pi  x}{2}\right) \Gamma (x) (-i\omega)^{-x}.
\eeq

At the saddle point, we have the effective action for the fluctuations of the $A_{ij}$ fields
\begin{align}
&S_{\mathrm{SP}}^A = t^2\frac{M}{N}\int d\tau d\tau^\pr \sum_{ij=1,i\le j}^N A_{ij}(\tau)\left[G(\tau-\tau^\pr)G(\tau^\pr-\tau)
-\delta(\tau-\tau^\pr)\int d\tau^{\pr\pr}G(\tau-\tau^{\pr\pr})G(\tau^{\pr\pr}-\tau)\right]A_{ij}(\tau^\pr)
\label{SASP}
\end{align}
Under the scaling $\tau\rightarrow b\tau$, we have $G(\tau)\rightarrow b^{x-1}G(\tau)$ from (\ref{gIR}), which then implies $A_{ij}(\tau)\rightarrow b^{-x}A_{ij}(\tau)$ from (\ref{SASP}). Corrections to (\ref{SASP}) coming from the expansion of $e^{i(A_{ij}(\tau)-A_{ij}(\tau^\pr))}$ beyond quadratic order in (\ref{S1}) are of the form $\int d\tau d\tau^\pr c_{n>2}(A_{ij}(\tau)-A_{ij}(\tau^\pr))^{n>2} G(\tau-\tau^\pr)G(\tau^\pr-\tau)$. The above scaling then implies that $c_{(n>2)}\rightarrow b^{(n-2)x}c_{(n>2)}$, so these terms are irrelevant, and their coefficients become small in the IR as $b\rightarrow0$, allowing us to ignore them.

\subsection{Deviations from half-filling}
\label{notahalf}

For $\mu\neq0$, the IR Green's function develops a spectral asymmetry, with $G(-\tau<0)=-e^{-2\pi\mathcal{E}}G(\tau>0)$ at $T=0$;
\beq
G(\tau>0) = -\frac{C(\mathcal{E})}{t^{1-x}\tau^{1-x}},~~G(\tau<0) = \frac{C(\mathcal{E})e^{-2\pi\mathcal{E}}}{t^{1-x}|\tau|^{1-x}}.
\label{gasym}
\eeq
The polarization $\Pi_{ij}(\tau)$ and the gauge field propagator however remain symmetric about $\tau=0$.  The real part of the self energy satisfies $\mathrm{Re}[\Sigma(i\omega_n\rightarrow0)]=\mu$, cancelling the chemical potential in the Green's function. The $t^2G(i\omega_n)$ term in the self-energy $\Sigma(i\omega_n)$ still cancels out as before. However, interestingly, the exponent $x$ of the power-law scaling depends on the asymmetry parameter $\mathcal{E}$ and is given by the solution to
\beq
\frac{(1/x-2)(\cosh(2\pi\mathcal{E})-\cos(\pi x))}{\tan(\pi x)\sin(\pi x)}=\frac{2M}{N}.
\label{xasym}
\eeq
This relation can be determined from the Dyson equation (\ref{Dyson1}) following Ref.~\cite{Sachdev1993}, and gives $x\rightarrow1/2$ as $\mathcal{E}\rightarrow\pm\infty$ regardless of $M/N$.  

\begin{figure}
\begin{center}
\includegraphics[height=2.0in]{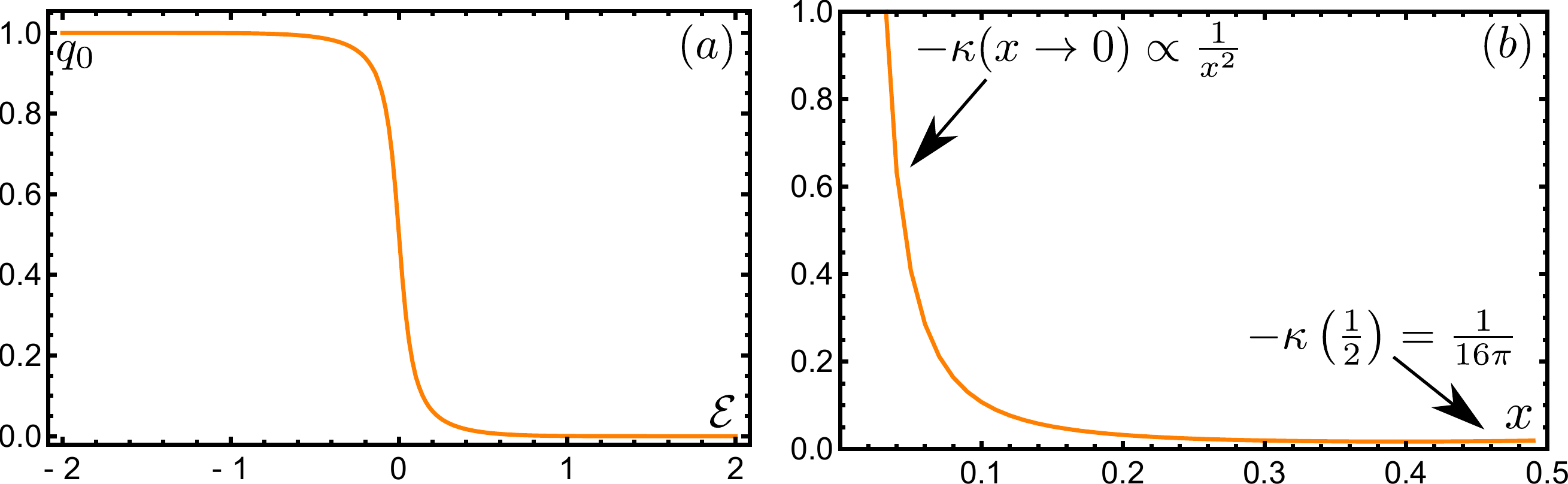} 
\end{center}
\caption{(a) Plot of the filling $q_0$ vs the asymmetry parameter $\mathcal{E}$ for $2M=N$. (b) Plot of the function $-\kappa(x)$ defined in (\ref{kappa}) vs the self-energy exponent $x$.}
\label{qefuncs}
\end{figure}

As in the SYK models~\cite{Sachdev2015,Sachdev2017}, the relationship between $\mathcal{E}$ and $\mu$ is nonuniversal, and depends on the values of UV details. However, following Ref.~\cite{Georges2001}, a universal relationship between the asymmetry parameter and the filling can be determined: The filling $q_0$ can be written as 
\beq
q_0 = i\int_{-\infty}^{\infty}\frac{dE}{2\pi}G^F(E)e^{iE0^+},
\eeq
where
\beq
G^F(E)\equiv\int_{-\infty}^{\infty}\frac{dE_1}{2\pi} \frac{\rho_f(E_1)}{E_1-E-i0^+\mathrm{sgn}(E_1)}
\eeq
is the Feynman Green's function, with $\rho_f(E_1)\equiv-2\mathrm{Im}[G^R(E_1)]$ the fermion spectral function. As in Ref.~\cite{Georges2001}, we have
\beq
q_0 = \frac{1}{\pi}(\mathrm{arg}[G^R(0^-)]-\mathrm{arg}[G^R(-\infty)]) + i\mathcal{P}\int_{-\infty}^{\infty}\frac{dE}{2\pi}\frac{\partial_E G^R(E)}{G^R(E)}e^{iE0^+} - i\mathcal{P}\int_{-\infty}^{\infty}\frac{dE}{2\pi}G^F(E)\partial_E \Sigma^F(E)e^{iE0^+},
\eeq
where $\mathcal{P}$ denotes the Cauchy principal value. Obtaining the low-energy forms of $G^R(E)$ and hence $\rho_f(E)$ from (\ref{gasym}), and using $G^R(E\rightarrow\pm\infty) = 1/(E+i0^+)$, this can then be written as
\beq
q_0 = 1 + \frac{x}{2} + \mathrm{arg}[-i\sin(\pi(x/2-i\mathcal{E}))e^{-i\pi x/2}] - i\mathcal{P}\int_{-\infty}^{\infty}\frac{dE}{2\pi}G^F(E)\partial_E \Sigma^F(E)e^{iE0^+}.
\eeq
The remaining integral needs to be computed carefully using the Dyson equation (\ref{Dyson1}) and the methods described in Appendix A of Ref.~\cite{Georges2001}. We obtain
\beq
q_0 = 1 + \frac{x}{2} + \mathrm{arg}[-i\sin(\pi(x/2-i\mathcal{E}))e^{-i\pi x/2}] - \frac{2N\kappa(x)\Gamma^2(x)\sin(2\pi x)\sinh(2\pi\mathcal{E})}{M\Gamma(2x-1)},
\label{luttinger}
\eeq 
where $\kappa(x)=\sum_{i=1}^6\mathcal{I}_i(x)$, with
\begin{align}
&\mathcal{I}_1 = -\frac{\cot(\pi x)\Gamma(1-x)\Gamma(2x)}{16\pi^2\Gamma(1+x)},~~\mathcal{I}_2 = -\frac{\csc(\pi x)\Gamma(2x)(\gamma_E+\psi(1+x))}{16\pi^2\Gamma(x)\Gamma(1+x)}, \nonumber 
\end{align}
\begin{align}
&\mathcal{I}_3 = -\frac{e^{i\pi x}}{16\pi^3}\int_{-\infty}^{-1}dY_1\int_{-\infty}^0 dY_2 \frac{Y_1^{-x}(-Y_2)^{2x-1}}{(Y_1+Y_2)(1+Y_1+Y_2)}, \nn
&\mathcal{I}_4 = \frac{1}{16\pi^3}\int_{-1}^0 dY_1 (-Y_1)^{-x} (Y_1+1)^{2 x-1} \left(-\frac{\, _2F_1\left(1,1-2 x;2-2 x;\frac{Y_1}{Y_1+1}\right)}{2x-1}-\ln (Y_1+1)+\gamma_E +\psi(1-2 x)\right), \nn
&\mathcal{I}_5 = -\frac{e^{i\pi x} x\Gamma(1-x)}{16\pi^3\Gamma(2-x)}\int_{-\infty}^0 dY_1\int_{-\infty}^0 dY_2\frac{\theta(-Y_1-Y_2-1)}{(Y_1+Y_2)^2} Y_1^{-x} (-Y_2)^{2 x-1}\, _2 F_1\left(1,1-x;2-x;-\frac{1}{Y_1+Y_2}\right),\nn
&\mathcal{I}_6 = -\frac{e^{i\pi x} x}{16\pi^3(1+x)}\int_{-\infty}^0 dY_1\int_{-\infty}^0 dY_2~\theta(Y_1+Y_2+1) Y_1^{-x} (-Y_2)^{2 x-1}\, _2 F_1\left(1,1+x;2+x;-(Y_1+Y_2)\right),
\label{kappa}
\end{align}
where $\psi$ is the digamma function, $\theta$ is the Heaviside step function, $\,_2F_1$ is a hypergeometric function~\cite{hypergeom}, and $\gamma_E$ is the Euler-Mascheroni constant. $\kappa(x\rightarrow1/2)=-1/(16\pi)$, and $\kappa(x\rightarrow0)\propto-1/x^2$ (see Fig.~\ref{qefuncs}b). Putting together (\ref{xasym}), (\ref{luttinger}) and (\ref{kappa}), we see that $q_0$ is a smooth function of $\mathcal{E}$ that decreases monotonically from $1$ to $0$ as $\mathcal{E}$ is swept from $-\infty$ to $\infty$ (see Fig.~\ref{qefuncs}a). This dependence of $q_0$ on $\mathcal{E}$ also agrees quantitatively  with that obtained from the numerical solutions of (\ref{Dyson1}), in which $q_0$ is given by $q_0=G(\tau=0^-)$ and $e^{-2\pi\mathcal{E}}=\mathrm{Im}[G^R(\omega=0^-)]/\mathrm{Im}[G^R(\omega=0^+)]$.

\subsection{Nonzero temperature}
\label{heated}

The regularized IR Dyson equations (\ref{Dyson1}) can be written in the time domain using two-time notation as (see Ref.~\cite{Sachdev2015})
\begin{align}
&-\int d\tau_3 G(\tau_1,\tau_3)\tilde{\Sigma}(\tau_3,\tau_2) = \delta(\tau_1,\tau_2),~~\tilde{\Sigma}(\tau_1,\tau_2) = t^2\nu G(\tau_1,\tau_2) + t^2G(\tau_1,\tau_2)D(\tau_2,\tau_1)\nn
&\int d\tau_3 D(\tau_1,\tau_3)\tilde{\Pi}(\tau_3,\tau_2) = \tilde{\delta}(\tau_1,\tau_2),~~\tilde{\Pi}(\tau_1,\tau_2) = 2t^2\frac{M}{N}\left(G(\tau_1,\tau_2)G(\tau_2,\tau_1)-\delta(\tau_1,\tau_2)\int d\tau_3 G(\tau_1,\tau_3)G(\tau_3,\tau_2)\right),
\label{IRrepar1}
\end{align}
where 
\beq
\nu = 1-\lim_{2\rightarrow1}[D(\tau_1,\tau_2) + D_{\mathrm{UV}}(\tau_1,\tau_2)],
\label{nudef}
\eeq
and $\tilde{\delta}(\tau_1,\tau_2)=\delta(\tau_1,\tau_2)-L^{-1}_\tau$, where $L_\tau$ is the length of the time domain ($L^{-1}_\tau=T$ at a finite temperature $T$). The chemical potential $\mu$ has been absorbed into $\Sigma$ to regularize it to $\tilde{\Sigma}$. We split the gauge field propagator into an IR piece $D$ and a UV piece $D_\mathrm{UV}$. The UV piece is not determined by (\ref{IRrepar1}), and is not sensitive to rescalings of $\tau$. The reason that the $\tilde{\delta}$ appears instead of just a $\delta$ is because the action (\ref{S2}) doesn't contain zero frequency modes of $A_{ij}$. As a result, $D$ here doesn't contain a zero frequency mode either, and consequently the pertinent delta function should be modified to remove its zero frequency mode. On a time domain of infinite size (such as at zero temperature), the zero frequency mode occupies a measure zero subspace, and then there is no difference between $\tilde{\delta}$ and $\delta$.  

The equations (\ref{IRrepar1}) are not invariant under a general set of reparametrizations with $\tau=f(\sigma)$ and an arbitrary function $h$,~\cite{Sachdev2015}
\begin{align}
&G(\tau_1,\tau_2) \rightarrow \frac{h(\sigma_1)/h(\sigma_2)}{(f'(\sigma_1)f'(\sigma_2))^a}G(\sigma_1,\sigma_2),~~\tilde{\Sigma}(\tau_1,\tau_2) \rightarrow \frac{h(\sigma_1)/h(\sigma_2)}{(f'(\sigma_1)f'(\sigma_2))^{1-a}}\tilde{\Sigma}(\sigma_1,\sigma_2), \nn
&D(\tau_1,\tau_2) \rightarrow (f'(\sigma_1)f'(\sigma_2))^{2a-1}D(\sigma_1,\sigma_2),~~\tilde{\Pi}(\tau_1,\tau_2) \rightarrow (f'(\sigma_1)f'(\sigma_2))^{-2a}\tilde{\Pi}(\sigma_1,\sigma_2),
\label{fullrepar}
\end{align}
because of the second term in the expression for $\tilde{\Pi}$, and additionally because $\nu$ and $L^{-1}_\tau$ can be nonzero. However, they can still be scale invariant under $\tau\rightarrow b\tau$ iff
\beq
G\rightarrow b^{-2a}G,~~\tilde{\Sigma}\rightarrow b^{2a-2}\tilde{\Sigma},~~D\rightarrow b^{4a-2}D,~~\tilde{\Pi}\rightarrow b^{-4a}\tilde{\Pi},~~\nu\rightarrow b^{4a-2}\nu.
\eeq
Note that $a$ is not determined by these equations, but we choose $2a=1-x$ due to the particular power-law scaling  of Sec.~\ref{gauge0} that is selected when the UV-complete equations are solved. 

\begin{figure}
\begin{center}
\includegraphics[height=2.2in]{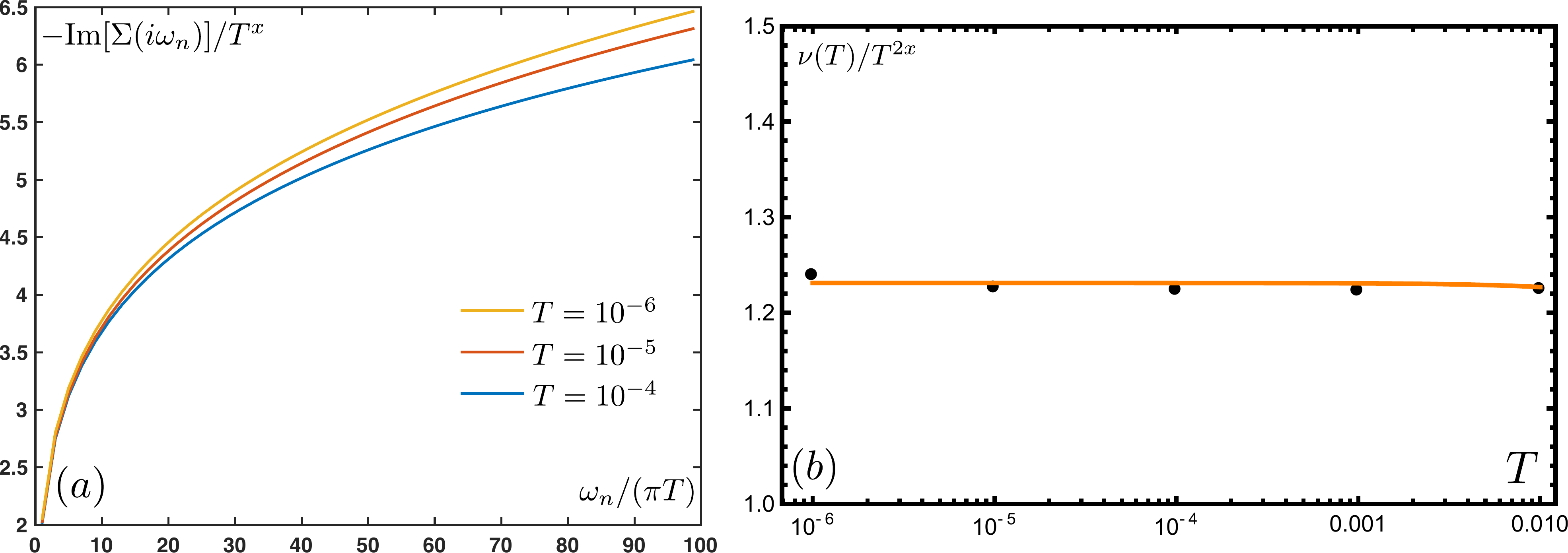} 
\end{center}
\caption{(a) Plot of the scaling form of the fermion self-energy $C^{-1}(0)t^{1-x}F_\Sigma(\omega_n/T,0)\equiv -\mathrm{Im}[\Sigma(i\omega_n)]/T^x$ vs $\omega_n/(\pi T)$ obtained by numerical solution of the imaginary-time Dyson equations for different values of $T$. For all curves, $t=g^2=1$, $2M=N$ and $\mu=0$ (corresponding to $x=0.230651$ from (\ref{expeq})). The curves collapse onto one another at low frequencies, confirming a universal low-energy scaling form. The deviations from universality at higher frequencies and temperatures occur because of the finite size of the critical window $\Lambda$ over which the low-energy solution is valid. (b) Plot of the quantity $\nu(T)/T^{2x}$ vs $T$ for the same values of parameters. This clearly shows $\nu(T)\propto T^{2x}$ over a large range of temperatures.}
\label{SEscaling}
\end{figure}

Now consider applying the scale transformation at a finite temperature. Since $\tau\in[0,1/T)$, this also scales $T\rightarrow T/b$, leaving $T\tau$ invariant. (\ref{IRrepar1}) is then compatible with a scaling solution (reverting back to one-time notation) $G(\tau)\propto T^{1-x}F_G(\tau T)$ (and corresponding expressions for $D$, $\tilde{\Sigma}$ and $\tilde{\Pi}$) iff $\nu\propto T^{2x}$. To check that we indeed get this behavior of $\nu$, we use the definition (\ref{nudef}) of $\nu$,  the fact that $D_\mathrm{UV}$ is not affected by rescalings of $T$ at low $T\ll \Lambda$, and the scaling form for $D(\tau)\propto T^{2x}F_D(\tau T)$, to obtain
\beq
\nu(T)-\nu(0) = \lim_{T\rightarrow0}\lim_{\tau\rightarrow0}[T^{2x}F_D(\tau T)]-\lim_{\tau\rightarrow0}T^{2x}F_D(\tau T),
\eeq
which gives $\nu(T)\propto T^{2x}$ when $\nu(0)=0$, which we already established in Sec.~\ref{gauge0}. Thus, the low-energy Dyson equations in the gauge field problem are fully consistent with a scaling solution at small finite temperatures. Our numerical solution confirms this (Fig.~\ref{SEscaling}a), and we also find $\nu(T)\sim T^{2x}$ numerically at small $T$ (Fig.~\ref{SEscaling}b). 

The IR fermion Green's function in the gauge field case does {\it not} have a conformally remapped form at $T\neq0$, as the equations (\ref{IRrepar1}) are not invariant under (\ref{fullrepar}) with $\tau=\tan(\pi T\sigma)/T$, but instead obeys
\beq
G(i\omega_n,T) = \frac{C(\mathcal{E})}{t^{1-x}T^x}F_G\left(\frac{\omega_n}{T},\mathcal{E}\right),~~F_G(y\rightarrow0,\mathcal{E})\propto y^0,~~F_G(y\rightarrow\infty,\mathcal{E}) \propto \frac{1}{y^x}.
\label{gIRT}
\eeq 
We can only compute the scaling function $F_G$ numerically. The self-energy also satisfies a low-energy scaling form $\mathrm{Im}[\Sigma(i\omega_n)]=-C^{-1}(\mathcal{E})t^{1-x}T^x F_\Sigma(\omega_n/T,\mathcal{E})$. However, the scaling function $F_\Sigma$ again differs from the conformal scaling function for the same exponent $x$  (corresponding to the self-energy $\Sigma_c(i\omega_n)=-1/G_c(i\omega_n)$ derived from the conformal Green's function $G_c(\tau)=-(C\pi^{1-x}/t^{1-x})(T/\sin(\pi T |\tau|))^{1-x}$), as can be seen by comparing universal ratios such as $n_{31} \equiv \mathrm{Im}[\Sigma(3i\pi T)]/\mathrm{Im}[\Sigma(i\pi T)]$ with their corresponding conformal values (see Table \ref{ratios}).

\begin{table}
\centering
\begin{tabular}{|l|l|l|l|l|}
\hline
Ratio    & $T=10^{-4}$ & $T=10^{-5}$ & $T=10^{-6}$ & Conformal \\ \hline
$n_{31}$ & 1.3680      & 1.3703      & 1.3709      & 1.2607    \\ \hline
$n_{53}$ & 1.1363      & 1.1382      & 1.1387      & 1.1224    \\ \hline
$n_{75}$ & 1.0845      & 1.0862      & 1.0865      & 1.0800    \\ \hline
$n_{97}$ & 1.0611      & 1.0626      & 1.0629      & 1.0594    \\ \hline
\end{tabular}
\caption{Comparision of numerical values of ratios $n_{ab}\equiv \mathrm{Im}[\Sigma(ia\pi T)]/\mathrm{Im}[\Sigma(ib\pi T)]$ in the gauge field problem at different temperatures with those derived from the conformal Green's function $G_c(\tau)=-(C\pi^{1-x}/t^{1-x})(T/\sin(\pi T |\tau|))^{1-x}$). As $T$ is reduced, these ratios converge to universal values that differ significantly from the conformal ones at low energies, which implies that the scaling function $F_\Sigma$ (or $F_G$) is not the conformal one, and that the local criticality in the gauge field model is different from the  SYK universality class. The values of other parameters used are the same as those used in Fig.~\ref{SEscaling}, but these universal low-energy ratios are insensitive to the values of $t$ and $g^2$ as $T\rightarrow0$ within numerical tolerances.}
\label{ratios}
\end{table}

\section{Thermodynamics}
\label{thermo}
In this section we describe the thermodynamic properties of the saddle-point solution described in the previous two sections at low temperatures. We specialize to the case of half-filling, with $\mu=0$. This allows for temperature derivatives of the free energy at a constant fermion density of half-filling to be the same as its temperature derivatives at constant zero chemical potential, which are easier to evaluate. We do not expect any of the qualitative features discussed here to be modified away from half-filling. 

The free energy can be written down from (\ref{S2intout}) evaluated at the saddle-point. It is
\begin{align}
&\mathcal{F} = -T \ln Z = MNT\sum_{\omega_n}\ln\left[\frac{i\omega_n}{i\omega_n-\Sigma_i(i\omega_n)}\right] -MNT\ln 2+t^2\frac{MN}{2} T\sum_{\omega_n} G^2(i\omega_n) -MNT\sum_{\omega_n}\Sigma(i\omega_n)G(i\omega_n) \nn
&+ \frac{N^2T}{4}\sum_{\Omega_m\neq0}\ln\left[\Pi(i\Omega_m)-\Pi(i\Omega_m=0)\right],
\label{fenerg}
\end{align}
where we added and subtracted the free fermion contribution, so that the frequency sum involving the logarithm converges and we may evaluate it numerically. The term on the second lime represents the gauge field contribution. Setting this aside for the moment, and numerically evaluating the fermion contribution using the saddle point of the UV complete action with the electric ``Maxwell" term (\ref{minUVc}), we obtain $\mathcal{F}_f/(MN) \approx -c_0(M/N) + T c_1(M/N)$ as $T\rightarrow0$. This implies that the fermion contribution to the specific heat $\mathcal{C}_f = -T\partial^2\mathcal{F}_f/\partial T^2$ vanishes at low temperatures and the fermions make a constant negative contribution $\mathcal{S}_f = -\partial \mathcal{F}_f/\partial T$ to the total entropy. Since the fermions and gauge fields are highly entangled, we of course need to add the gauge field contribution to obtain the full physical free energy and associated thermodynamic quantities. We can write 
\beq
\mathcal{F}_A = \frac{N^2T}{4}\sum_{\Omega_m\neq0} \ln\left[\frac{\Pi(i\Omega_m)-\Pi(i\Omega_m=0)}{\Pi(i\Omega_m,T=0)-\Pi(i\Omega_m=0,T=0)}\right] + \frac{N^2T}{4}\sum_{\Omega_m\neq0} \ln\left[\Pi(i\Omega_m,T=0)-\Pi(i\Omega_m=0,T=0)\right],
\label{fale}
\eeq
where we have added and subtracted a term that is evaluated using the zero temperature functional form of $\Pi$, but evaluated at the Matsubara frequencies corresponding to a particular temperature. Since $\Pi(i\Omega_m)-\Pi(i\Omega_m=0)$ obeys a quantum-critical scaling form, the first term becomes $\mathcal{F}_A^{(1)}/(MN)=-\frac{NT}{4M}\sum_{m\neq0}\ln[(2\pi m)^{1-2x}F_D(2\pi m)]$. We find numerically that $F_D(2\pi m)= 1/(2\pi m)^{1-2x}+\mathcal{O}(1/m^2)$ for $m\gg 1$ in the scaling limit, so this sum converges at large $m$, and leads to $\mathcal{F}^{(1)}_A/(MN)\sim -T$, which doesn't contribute to the specific heat and provides a constant contribution to the entropy at low temperatures.

The second term of (\ref{fale}) can be computed by zeta-function regularization using the result (\ref{pIR}):
\begin{align}
&\frac{\mathcal{F}_A^{(2)}}{MN} = \frac{NT}{4M}\sum_{m\neq0}\ln\left[-4\frac{M}{N}C^2 t^{2x}\sin (\pi  x) \Gamma (2 x-1) \left|2\pi m\right|^{1-2x}T^{1-2x}\right] \nn
&=-\frac{NT}{4M}\ln\left[-4\frac{M}{N}C^2 t^{2x}\sin (\pi  x) \Gamma (2 x-1)T^{1-2x}\right].
\end{align}
This produces the dominant contribution to the low-temperature specific heat, 
\beq
\frac{\mathcal{C}}{MN}\approx\frac{\mathcal{C}_A^{(2)}}{MN}=-T\frac{\partial^2\mathcal{F}_A^{(2)}}{\partial T^2} = \frac{(1-2x)N}{4M},
\eeq
which is positive and extensive. In the limit of $M/N\rightarrow\infty$, where $x\rightarrow0$ and the non-Fermi liquid solution turns into a noninteracting random matrix solution, this large contribution to the specific heat vanishes as it should, and in the opposite limit of  $M/N\rightarrow0$, where $x\rightarrow1/2$, it blows up as $1/\sqrt{M/N}$, as can be seen by applying (\ref{expeq}).

The free energy contribution $\mathcal{F}_A^{(2)}$ also leads to the dominant contribution to the low-temperature entropy $\mathcal{S}_A^{(2)}/(MN)=-\partial\mathcal{F}_A^{(2)}/\partial T \propto (1-2x)(N/(4M))\ln T$. This is negative at low $T$, which indicates that our theory is incomplete: extra degrees of freedom must be present in a physical theory in order to offset this entropy. The reason this happens is that our theory is missing all information about the zero-frequency modes of the $A_{ij}$. In any sensible electromagnetic lattice gauge theory, these modes will contribute to physical static magnetic field configurations that cost energy: Exciting a single link $A_{ij}$ will lead to nonzero magnetic fluxes through all plaquettes containing that link, and a magnetic ``Maxwell" term acting on these fluxes will contribute to the action, even if they are static. However such terms are not generated in our theory by integrating out the fermions in the large-$N$,$M$ limits. In order to generate these terms we need to appeal to some heavy degrees of freedom that couple to the gauge fields in such a way that integrating out these degrees of freedom will produce magnetic ``Maxwell" terms. 

Assuming this is the case, we write down the simplest possible gauge and time-reversal invariant magnetic ``Maxwell" action that is appropriate for an all-to-all interacting theory without any spatial structure. It is
\beq
S_B = \frac{m^2_B}{2(N-2)}\int d\tau \sum_{\triangle ijk}(A_{ij}(\tau)+A_{jk}(\tau)+A_{ki}(\tau))^2,
\eeq
where the sum runs over all possible unique triangles. The kernel of this quadratic action has $(N-1)(N-2)/2$ degenerate eigenvectors with eigenvalue $m_B^2(1+2/(N-2))$ and $N-1$ degenerate eigenvectors with eigenvalue $0$. The zero-eigenvalued eigenvectors are all pure gauge and can each be gauge-transformed to the configuration $A_{ij}=0$; they correspond to the state in which the flux through all triangles is zero, and thus do not contribute anything to the free energy. In the large-$N$ limit, the thermodynamic fraction of modes residing on a single links $A_{ij}$ have negligible overlap with the zero-eigenvalued eigenvectors. This permits the approximation, exact in the infinite-$N$ limit
\beq
S_B \approx \frac{m^2_B}{2}\int d\tau \sum_{ij=1,i\leq j}^N A_{ij}^2(\tau).
\eeq
We assume that $m^2_B$ is much smaller than $-4(M/N)C^2\sin(\pi x)\Gamma(2x-1)t^{2x}T^{1-2x}$. Then including this term just adds 
\beq
\mathcal{F}_A^{(3)} = \frac{N^2}{4}T\ln m^2_B
\eeq
to (\ref{fenerg}). This term doesn't contribute to the specific heat, but offsets the leading contribution to the entropy to a large positive value $\mathcal{S}_A^{(2)+(3)}/(MN) = (N/(4M))\ln (-4(M/N)C^2\sin(\pi x)\Gamma(2x-1)t^{2x}T^{1-2x}/m^2_B)$.

For $-4(M/N)C^2\sin(\pi x)\Gamma(2x-1)t^{2x}T^{1-2x}\ll m_B^2$, the fermions effectively end up coupling to gapped bosonic modes. The low-energy Dyson equation then reads
\begin{align}
&\Sigma(i\omega_n)=t^2G(i\omega_n)\left[1-T\sum_{\Omega_m\neq0}\frac{1}{m_B^2+\Pi(i\Omega_m)-\Pi(i\Omega_m=0)}\right] +t^2T\sum_{\Omega_m\neq0}\frac{G(i\omega_n+i\Omega_m)}{m_B^2+\Pi(i\Omega_m)-\Pi(i\Omega_m=0)}, \nn
&\Pi(i\Omega_m) = 2t^2T\frac{M}{N}\sum_{\omega_n}G(i\omega_n)G(i\omega_n+i\Omega_m),~~G(i\omega_n)  = \frac{1}{i\omega_n+\mu-\Sigma(i\omega_n)}.
\label{Dysongapped}
\end{align}
The term in square brackets no longer cancels at $T=0$, as increasing the denominator of the boson propagator by adding a mass makes its value smaller than the zero-mass case. This leaves behind a $\nu t^2 G(i\omega_n)$ term in $\Sigma(i\omega_n)$, leading to a renormalized random-matrix solution at the lowest energies. The second term in the first line of (\ref{Dysongapped}) vanishes at small external frequencies $\omega_n$, as $G(i\Omega_m)$ is odd in $\Omega_m$ and the denominator is a constant at low frequencies (for nonzero chemical potential, this sum just produces a constant that is absorbed by $\mu$). These points can be easily verified by numerically solving the UV-completed version of (\ref{Dysongapped}) using the \verb|MATLAB| code \verb|gd.m|~\cite{gd}. The lowest energy state then has a vanishing entropy and specific heat. Henceforth, we shall assume that we are only interested in energy scales larger than the small $m^2_B$, treating it as an IR regulator much smaller than $T$, and focus on the non-Fermi liquid.

We also checked numerically that the compressibility $MN\partial q_0/\partial\mu|_T$, where $q_0=G(\tau=0^-)$ asymptotes to a nonzero constant as $T\rightarrow0$. This justifies our rationale of ignoring the time components $A^0_i$ of the gauge fields in the IR, as their propagators are screened by this compressibility. 

Finally, from the point of view of the magnetic ``Maxwell" terms, the model behaves like a U(1) gauge theory in a large ($\mathcal{O}(N)$) number of dimensions. Possible magnetic monopoles arising due to the compactness of the U(1) gauge group then source nonzero fluxes through a large number of plaquettes, leading to $\mathcal{O}(N)$ increases in the free energy through the magnetic Maxwell terms, while not coupling to the fermions by virtue of being a static background. Thus, the configuration in which no monopoles exist should be a stable saddle-point, and monopole operators are irrelevant.    

\section{Transport}
\label{transport}

In order to consider transport properties of this model, we need to make appropriate modifications. First, we need some spatial structure. This can be achieved by defining the clusters indexed by $i,j$ to lie on the sites of an $N$-dimensional hypercubic lattice, with each cluster then having $2N$ neighbors. The fermions hop between nearest neighbor clusters, coupling to gauge fields $A_{ij}$ that live on the bonds of the lattice. Second, for an external probe gauge field to drive a current, it must couple to a different charge from the one that the internal gauge fields $A_{ij}$ couple to: If they coupled to the same charge, then turning on the probe field only amounts to shifting the values of $A_{ij}$, and the path integral over $A_{ij}$ trivially absorbs these shifts, rendering the partition function immune to the probe field. If we view the fermions as chargons arising from fractionalization in an ACL, we can divide the flavors indexed by $\alpha,\beta$ into equal fractions of two species that couple to the internal gauge field with opposite charges, but which couple to the external probe gauge field with equal charges, which is a single-axis version of the SU(2) case discussed in Ref.~\cite{SachdevReview2018}. Then, our modified version of (\ref{H}) reads
\begin{align}
&\mathcal{H}^\pr  = -\frac{1}{(2MN)^{1/2}}\sum_{\langle ij \rangle}\sum_{\alpha\beta=1}^{M} \sum_{ss^\pr=\pm} \left[t_{ij}^{\alpha \beta }f^\dg_{i\alpha s}e^{iA_{ij}\sigma^z_{ss^\pr}} f_{j\beta s^\pr}+(2MN)^{1/2}\mu \delta_{ij}^{\alpha \beta}\delta_{ss^\pr} f^\dg_{i\alpha s}f_{i\alpha s}\right], \nn
&\ll t_{ij}^{\alpha \beta} t_{ji}^{\beta \alpha} \gg =\ll |t_{ij}^{\alpha \beta}|^2\gg = t^2.
\label{Hpr}
\end{align}
This has a U(1) gauge invariance under $f_{i\alpha s}(\tau)\rightarrow \sum_{s^\prime=\pm}e^{i\theta_i(\tau)\sigma^z_{ss^\prime}}f_{i\alpha s^\prime}(\tau)$ and $A_{ij}(\tau)\rightarrow A_{ij}(\tau)+\theta_i(\tau)-\theta_j(\tau)$.

Performing the same manipulations as before, we obtain
\begin{align}
&S^\pr = \int d\tau \sum_{i}\sum_{\alpha=1}^{M}\sum_{s=\pm} f^\dg_{i\alpha s}(\tau)(\partial_\tau+isA^0_i(\tau)+\mu)f_{i\alpha s}(\tau) \nn
&+t^2\frac{M}{2N}\int d\tau d\tau^\pr \sum_{\langle ij\rangle} \sum_{ss^\pr=\pm} \left[\left(1-\frac{1}{2}A_{ij}^2(\tau)-\frac{1}{2}A_{ij}^2(\tau^\pr)+A_{ij}(\tau)A_{ij}(\tau^\pr)\right)\delta_{ss^\pr}+i(A_{ij}(\tau)-A_{ij}(\tau^\pr))\sigma^z_{ss^\pr}\right] \nn
&\times G_{j s^\pr}(\tau-\tau^\pr)G_{i s}(\tau^\pr-\tau) -M\int d\tau d\tau^\pr \sum_{i}\sum_{s=\pm}\Sigma_{is}(\tau-\tau^\pr)\left[G_{is}(\tau^\pr-\tau)-\frac{1}{M}\sum_{\alpha=1}^Mf_{i\alpha s}(\tau^\pr)f^\dg_{i\alpha s}(\tau)\right],
\label{Spr}
\end{align} 
as before, the time integrations kill the term proportional to $\sigma^z$ in the second line of the above. This action then leads to a saddle-point symmetric in $s$ described by (\ref{Dyson1}), with the IR solution (\ref{gIR}). Similar arguments for invariance under gauge-fixing at large-$N$ and stability of the saddle-point as before apply. 

We now perturb the action (\ref{Spr}) with a diagonal probe field, so that $A_{ij}(\tau)\sigma^z_{ss^\pr}\rightarrow A_{ij}(\tau)\sigma^z_{ss^\pr}+\Xi_{ij}(\tau)\delta_{ss^\pr}$ where $\Xi_{ij}(\tau)=\delta_{j,i+\hat{x}}\Xi(\tau)$, which corresponds to applying an electric field $\mathbf{E}=-(d\Xi(\tau)/d\tau)\hat{x}$ in the $\hat{x}$ direction. The perturbed action reads
\begin{align}
&S^\pr_\Xi =  -M\sum_{i}\sum_{s=\pm} \mathrm{Tr}\ln[\partial_\tau+\mu\delta(\tau,\tau^\pr)-\Sigma_{is}(\tau,\tau^\pr)] \nn
&+t^2\frac{M}{2N}\int d\tau d\tau^\pr \sum_{\langle ij\rangle} \sum_{ss^\pr=\pm} \left[\left(1-\frac{1}{2}(A_{ij}(\tau)-A_{ij}(\tau^\pr))^2\right)\delta_{ss^\pr}+i(A_{ij}(\tau)-A_{ij}(\tau^\pr))\sigma^z_{ss^\pr}\right]G_{j s^\pr}(\tau,\tau^\pr)G_{i s}(\tau^\pr,\tau) \nn
&+t^2\frac{M}{2N}\int d\tau d\tau^\pr \sum_{\langle ij\rangle} \sum_{s=\pm} \left[i(\Xi_{ij}(\tau)-\Xi_{ij}(\tau^\pr))-\frac{1}{2}(\Xi_{ij}(\tau)-\Xi_{ij}(\tau^\pr))^2\right]G_{j s}(\tau,\tau^\pr)G_{i s}(\tau^\pr,\tau) \nn
&-t^2\frac{M}{2N}\int d\tau d\tau^\pr \sum_{\langle ij\rangle} \sum_{ss^\pr=\pm} (A_{ij}(\tau)-A_{ij}(\tau^\pr))(\Xi_{ij}(\tau)-\Xi_{ij}(\tau^\pr))\sigma^z_{ss^\pr}G_{j s^\pr}(\tau,\tau^\pr)G_{i s}(\tau^\pr,\tau),
\label{Sprpert}
\end{align} 
where we integrated out the fermions and neglected the $A^0_i$ as before. With the perturbed partition function $Z^\pr_\Xi=\int \mathcal{D}A\mathcal{D}G\mathcal{D}\Sigma~e^{-S^\pr_\Xi[A,G,\Sigma]}$, we then obtain the current-current correlator 
\beq
\langle J_x(\tau)J_x(\tau^\pr) \rangle = \frac{1}{Z^\pr_{\Xi=0}}\frac{\delta^2Z^\pr_\Xi}{\delta\Xi(\tau)\delta\Xi(\tau^\pr)}\Bigg|_{\Xi=0}=\int \mathcal{D}A\mathcal{D}G\mathcal{D}\Sigma~\frac{e^{-S^\pr_{\Xi=0}[A,G,\Sigma]}}{Z^\pr_{\Xi=0}}\left(\frac{\delta S^\pr_{\Xi}[A,G,\Sigma]}{\delta \Xi(\tau)}\frac{\delta S^\pr_{\Xi}[A,G,\Sigma]}{\delta \Xi(\tau^\pr)}-\frac{\delta^2 S^\pr_{\Xi}[A,G,\Sigma]}{\delta\Xi(\tau)\delta \Xi(\tau^\pr)}\right)\Bigg|_{\Xi=0}.
\eeq
The only term that survives after integrating out the fields (which makes $G$ and $\Sigma$ take their saddle-point values) is 
\beq
\langle J_x(\tau)J_x(\tau^\pr) \rangle = -Vt^2\frac{M}{N}\left[G(\tau-\tau^\pr)G(\tau^\pr-\tau)-\delta(\tau-\tau^\pr)\int d\tau^{\pr\pr}G(\tau-\tau^{\pr\pr})G(\tau^{\pr\pr}-\tau)\right],
\label{JJ}
\eeq
where $V$ is the system volume (number of sites in the hypercubic lattice). The right-hand-side of (\ref{JJ}) automatically contains the sum of the paramagnetic and diamagnetic terms. 

This gives rise to the DC conductivity, employing the scaling forms derived in Sec.~\ref{heated},
\beq
\sigma_{xx}^{\mathrm{DC}}=-\frac{1}{V}\lim_{\mathrm{\Omega_m\rightarrow0}}\frac{\langle J_x J_x\rangle(i\Omega_m)}{\Omega_m}\sim \frac{M}{N}\left(\frac{t}{T}\right)^{2x},
\label{sigDC}
\eeq
and the optical conductivity
\beq
\sigma_{xx}(\Omega\gg T)= -2(M/N)C^2 \sin (\pi  x) \Gamma (2 x-1) \left(\frac{it}{\Omega}\right)^{2x}.
\label{sigAC}
\eeq
As discussed in Sec.~\ref{model}, since the saddle-point value of $G$ is gauge-independent at leading order in large-$N$, this answer for the conductivity is correctly gauge-invariant at leading order in large-$N$. Since the critical solution (\ref{gIRT}) is in general valid only for $T\ll t$, the DC conductivity (\ref{sigDC}) is never parametrically in a bad-metallic regime of $\sigma^{\mathrm{DC}}\ll 1$ within the energy window of validity of the non-Fermi liquid solution. 

\section{Discussion}
\label{discuss}

We have constructed a model of a disordered non-Fermi liquid phase of fermions at a finite density coupled to gapless fluctuating U(1) gauge fields, in a solvable large-$N$ limit. In this non-Fermi liquid phase, both the fermion and photon Green's functions are gapless, and decay as power-laws of time at long times. The power-law exponents are continuously tunable within a finite range, and, interestingly, depend upon the filling fraction of the fermions. 

A special feature of our model is that the non-Fermi liquid phase arises under the combined effect of hopping and interaction terms, in contrast to the purely interacting SYK models. In the SYK models, the addition of quadratic hopping terms results in a weakly-interacting Fermi liquid solution in the infrared~\cite{Balents2017}. However, unlike the SYK models, in which the interaction between the fermions is instantaneous in the large-$N$ limit, the interaction between fermions in our model is retarded, mediated by gapless bosonic modes with singular propagators at low energies, leading to non-Fermi liquid behavior even in the presence of hopping terms~\cite{Lee2018}.  

Our model only possesses scale invariance in the infrared, and not the much more comprehensive time reparametrization invariance of the SYK models. At nonzero temperatures, this lack of time reparametrization symmetry in our model results in different finite temperature fermion Green's functions from the conformal ones that appear in the generalized set of SYK$_q$ models with $1<q/2<2$-body interactions~\cite{Maldacena2016,Sachdev2017}. Consequently, we do not expect our model to have as
direct a holographic connection to AdS$_2$ gravity as the SYK models, or to display maximal chaos \cite{SS10,kitaev2015talk,Sachdev2015,Maldacena2016,KitaevSuh}. However, due to the quantum-critical scaling of the Green's functions, we still expect the Lyapunov exponent for many-body quantum chaos to be an $\mathcal{O}(1)$ number times $k_BT/\hbar$, similar to other models of fermions strongly coupled to fluctuating gauge fields~\cite{Patel2017}. 

The dynamic photon modes cause our model to have a much larger Hilbert space than the SYK models, which only have fermions. This appears to allow for a finer spacing of the low-lying many-body energy levels than in the SYK models (which have a level spacing of $\sim e^{-N}$~\cite{Fu2016}), leading to parametrically larger values of entropy and specific heat at low temperatures, that are dominated by contributions from the photon modes.    

We can view our model as a toy model of an ACL~\cite{SachdevReview2018,Kaul2008}, which is a candidate for the strange metal regime of the cuprate superconductors. This is an effective theory in which electrons are fractionalized into gapless fermionic chargons which carry their charge (but not spin), and gapped bosonic spinons that do not affect the low-energy fluctuations of the chargons. By defining our model on an $N$-dimensional hypercubic lattice, we obtain non-Fermi liquid charge transport properties, with a sub-linear power-law-in-temperature resistivity.  The exponent of the power-law is continuously tunable as a function of the filling, and can approach linear-in-temperature for certain parameter ranges. This non-Fermi liquid has a `large Fermi surface', {\em i.e.\/} all $M$ flavors of fermions are active and contribute to transport. This is in contrast to the SYK/Kondo-lattice models of non-Fermi liquids proposed in Refs.~\cite{Patel2018,Chowdhury2018}, where only the itinerant fermions contribute to transport. 

For future work, it would be interesting to see if some of the strategies employed here can be extended to construct solvable models of fermions at finite density and with quenched disorder interacting with gauge fields in $2+1$ dimensions. Such models would of course be more realistic candidates for describing the phase diagram of the cuprates. It would also be interesting, if possible, to consider Higgs transitions out of ACLs in such models into weakly interacting `pseudogap' phases with a reduced number of active fermions~\cite{SachdevReview2018,Chowdhury2015}, along the lines of the analysis in Appendix~\ref{higgs}. 

\section*{Acknowledgements}

This research was supported by the NSF under Grant DMR-1664842. A. A. P.  was supported by a Harvard-GSAS Merit Fellowship. Research at Perimeter Institute is supported by the Government of Canada through Industry Canada and by the Province of Ontario through the Ministry of Research and Innovation. S. S. also acknowledges support from Cenovus Energy at Perimeter Institute.

\appendix

\section{Higgs transition from the U(1) ACL to a $\mathbb{Z}_2$ ACL}
\label{higgs}

We consider a Higgs transition that breaks the U(1) gauge invariance down to $\mathbb{Z}_2$ in the ACL of Sec.~\ref{transport}. This is expected to be a toy model
of the optimal doping transition in the cuprates without a symmetry-breaking order parameter, from the overdoped to the underdoped side \cite{SS09,Chowdhury2015,SSNambu,Scheurer18}. We modify the fermion-gauge field hamiltonian to
\beq
\mathcal{H}_1^{\pr\pr}  = -\frac{1}{(2MN)^{1/2}}\sum_{\langle ij \rangle}\sum_{\alpha\beta=1}^{M} \sum_{s=\pm} \left[t_{ijs}^{\alpha \beta }f^\dg_{i\alpha s}e^{i s A_{ij}} f_{j\beta s}+(2MN)^{1/2}\mu \delta_{ij}^{\alpha \beta} f^\dg_{i\alpha s}f_{i\alpha s}\right],~~\ll t_{ijs}^{\alpha \beta} t_{jis}^{\beta \alpha} \gg =\ll |t_{ijs}^{\alpha \beta}|^2\gg = t^2.
\label{Hpr2}
\eeq
We have now broken the $+\leftrightarrow-$ pseudospin symmetry since the hopping matrix elements $t^{\alpha\beta}_{ijs}$ are uncorrelated between $s=\pm$. However, this  symmetry is restored upon disorder-average as the variances of the $t^{\alpha\beta}_{ijs}$ are the same for $s=\pm$. This will allow us to easily write down saddle-point equations in the higgsed phase, as the 4-Fermi term produced by disorder-averaging will not have decompositions in the $\langle f^\dg_+f_-\rangle$ channel that would prevent its decomposition exclusively into the $G_i$'s. As before, the addition of Maxwell terms and time components for the gauge fields to $\mathcal{H}_1^{\pr\pr}$ is implied.

Now we add complex scalar Higgs fields $H_i$ defined on each site $i$ of the $N$-dimensional hypercube into the mix. These fields are charge $2$ under the U(1) gauge field, with $H_i\rightarrow H_i e^{2i\theta_i}$ under the U(1) gauge transformation.
\beq
\mathcal{H}_2^{\pr\pr}  = \sum_i\left[Mr|H_i|^2+g_H\left(H_i\sum_{\alpha=1}^Mf^\dg_{i\alpha+}f_{i\alpha-}+\mathrm{h.c.}\right)\right]-\frac{t_H}{2}\sum_{\langle ij\rangle}\left[H_i^\ast H_j e^{2iA_{ij}}+\mathrm{h.c.}\right].
\eeq
The addition of coupling to time components of the gauge fields to $\mathcal{H}_2^{\pr\pr}$ is implied. The couplings of the Higgs fields to the fermions are {\it non-random}, but a large-$M,N$ saddle-point can still be defined as was done in Ref.~\cite{PatelKim2018}, which had non-random couplings to a superconducting order parameter. To see this, we disorder-average the action of $\mathcal{H}_1^{\pr\pr}+\mathcal{H}_2^{\pr\pr}$ and then expand the exponentials to quadratic order as before (ignoring the screened time components of the gauge fields),
\begin{align}
&S^{\pr\pr} = \int d\tau \sum_{i}\sum_{\alpha=1}^{M}\left[\sum_{s=\pm} f^\dg_{i\alpha s}(\tau)(\partial_\tau+\mu)f_{i\alpha s}(\tau)+g_H\left(f^\dg_{i\alpha +}(\tau)H_i(\tau)f_{i\alpha -}(\tau)+\mathrm{h.c.}\right)\right] \nn
&+t^2\frac{M}{2N}\int d\tau d\tau^\pr \sum_{\langle ij\rangle} \sum_{s=\pm} \left[\left(1-\frac{1}{2}A_{ij}^2(\tau)-\frac{1}{2}A_{ij}^2(\tau^\pr)+A_{ij}(\tau)A_{ij}(\tau^\pr)\right)+is(A_{ij}(\tau)-A_{ij}(\tau^\pr))\right] \nn
&\times G_{js}(\tau,\tau^\pr)G_{is}(\tau^\pr,\tau) -M\int d\tau d\tau^\pr \sum_{i}\sum_{s=\pm}\Sigma_{is}(\tau,\tau^\pr)\left[G_{is}(\tau^\pr,\tau)-\frac{1}{M}\sum_{\alpha=1}^Mf_{i\alpha s}(\tau^\pr)f^\dg_{i\alpha s}(\tau)\right] \nn
&+M\int d\tau\sum_i\left[|\partial_\tau H_i(\tau)|^2+r|H_i(\tau)|^2\right]-\frac{t_H}{2}\int d\tau\sum_{\langle ij\rangle}\left[H_i^\ast(\tau)\left(1+2iA_{ij}(\tau)-2A_{ij}^2(\tau)\right)H_j(\tau)+\mathrm{h.c.}\right].
\label{SHiggs}
\end{align} 
We now integrate out the fermions and gauge fields
\begin{align}
&S^{\pr\pr} = -M\sum_{i}\mathrm{Tr}\ln\left(\begin{matrix}
\partial_\tau+\mu\delta(\tau,\tau^\pr) -\Sigma_{i+}(\tau,\tau^\pr) & g_H H_i(\tau)\delta(\tau,\tau^\pr) \\
g_H H_i^\ast(\tau)\delta(\tau,\tau^\pr) & \partial_\tau+\mu\delta(\tau,\tau^\pr) -\Sigma_{i-}(\tau,\tau^\pr) 
\end{matrix}\right) +M\int d\tau\sum_i\left[|\partial_\tau H_i(\tau)|^2+r|H_i(\tau)|^2\right] \nn
&+ \frac{1}{2}\sum_{\langle ij\rangle} \mathrm{Tr}\ln\left[-\frac{\partial_\tau^2}{g^2}+\tilde{\Pi}_{ij}(\tau,\tau^\pr)+2t_H(H_i^\ast(\tau)H_j(\tau)+\mathrm{h.c.})\delta(\tau,\tau^\pr)\right] -\frac{t_H}{2}\int d\tau \sum_{\langle ij\rangle} \left[H_i^\ast(\tau)H_j(\tau)+\mathrm{h.c.}\right] \nn
&+t^2\frac{M}{2N}\int d\tau d\tau^\pr \sum_{\langle ij\rangle} \sum_{s=\pm} G_{js}(\tau,\tau^\pr)G_{is}(\tau^\pr,\tau) -M\int d\tau d\tau^\pr \sum_{i}\sum_{s=\pm}\Sigma_{is}(\tau,\tau^\pr)G_{is}(\tau^\pr,\tau), \nn
&\tilde{\Pi}_{ij}(\tau,\tau^\pr)=t^2\frac{M}{N}\sum_{s=\pm}\left[G_{is}(\tau^\pr,\tau)G_{js}(\tau,\tau^\pr)-\frac{1}{2}\delta(\tau,\tau^\pr)\int d\tau^{\pr\pr}\left(G_{is}(\tau,\tau^{\pr\pr})G_{js}(\tau^{\pr\pr},\tau)+G_{is}(\tau^{\pr\pr},\tau)G_{js}(\tau,\tau^{\pr\pr})\right)\right].
\label{SHiggsio}
\end{align} 
where we threw out some terms that do not contribute to first-order variations at the saddle-point we will obtain. In addition to the saddle-point for $G_{is}$ and $\Sigma_{is}$, this action also has a saddle-point for $H_i$. Fluctuations of $H_i$ about this saddle point are suppressed by the large-$M$ limit. The combined saddle-point equations obtained by varying $G_{is}$, $\Sigma_{is}$ and $H_i$ about an $i,s$-uniform solution with constant $|H(\tau)|=|H|$ are
\begin{align}
&\Sigma(i\omega_n)=t^2G(i\omega_n) +t^2T\int\frac{d\Omega_m}{2\pi}\frac{G(i\omega_n+i\Omega_m)-G(i\omega_n)}{\Omega_m^2/g^2+\tilde{\Pi}(i\Omega_m)+4t_H |H|^2},~~G(i\omega_n)=\frac{i\omega_n+\mu-\Sigma(i\omega_n)}{(i\omega_n+\mu-\Sigma(i\omega_n))^2-g_H^2 |H|^2}, \nn
&H\left[r-\frac{N}{M}t_H+\int\frac{d\omega_n}{2\pi}\frac{g_H^2}{(i\omega_n+\mu-\Sigma(i\omega_n))^2-g_H^2 |H|^2}+\frac{2N}{M}\int\frac{d\Omega_m}{2\pi}\frac{t_H}{\Omega_m^2/g^2+\tilde{\Pi}(i\Omega_m)+4t_H |H|^2}\right]=0, \nn
&\tilde{\Pi}(i\Omega_m) = 2t^2\frac{M}{N}\int \frac{d\omega_n}{2\pi}G(i\omega_n)(G(i\omega_n+i\Omega_m)-G(i\omega_n)).
\label{Dysonhiggs}
\end{align}

Saddle-points for which $H$ is static in time with a spatially uniform magnitude but spatially varying phase are gauge-equivalent to the uniform solution, and yield the same fermion Green's function. For $r$ between $Nt_H/M$ and 
\beq
r_c \equiv -\frac{N}{M}t_H-\int\frac{d\omega_n}{2\pi}\frac{g_H^2}{(i\omega_n+\mu-\Sigma(i\omega_n))^2}\Bigg|_{H=0},
\eeq
the equations (\ref{Dysonhiggs}) have a solution with a Higgs condensate $|H|\neq0$, with $|H|$ vanishing as $r\rightarrow r_c$. In this higgsed phase, the only remaining gauge redundancy is a $\mathbb{Z}_2$ gauge transformation of $f\rightarrow-f$. The condensate renders the low-energy fluctuations of the gauge fields non-singular, which causes the low-energy fermion Green's function and self-energy to take on a random-matrix form $G(i\omega_n),\Sigma(i\omega_n)\sim i\mathrm{sgn}(\omega_n)$ for $g_H |H|\ll t$. The reasoning behind this is the same as that for the solution of (\ref{Dysongapped}), and the random-matrix like solution at low energies can easily be verified by solving (\ref{Dysonhiggs}) numerically using the \verb|MATLAB| code \verb|gdHiggs.m|~\cite{gdHiggs}. Relative to the non-Fermi liquid U(1) ACL phase, the low-energy fermion density of states $\sim\mathrm{Im}[G^R(\omega)]$ is thus depleted, akin to a `pseudogap' phase. Furthermore, the resistivity in the higgsed phase, following from (\ref{JJ}), becomes Fermi-liquid like, with $\rho^\mathrm{DC}_{xx}\sim \rho_0 + \rho_1 T^2$.

\begin{figure}
\begin{center}
\includegraphics[height=1.85in]{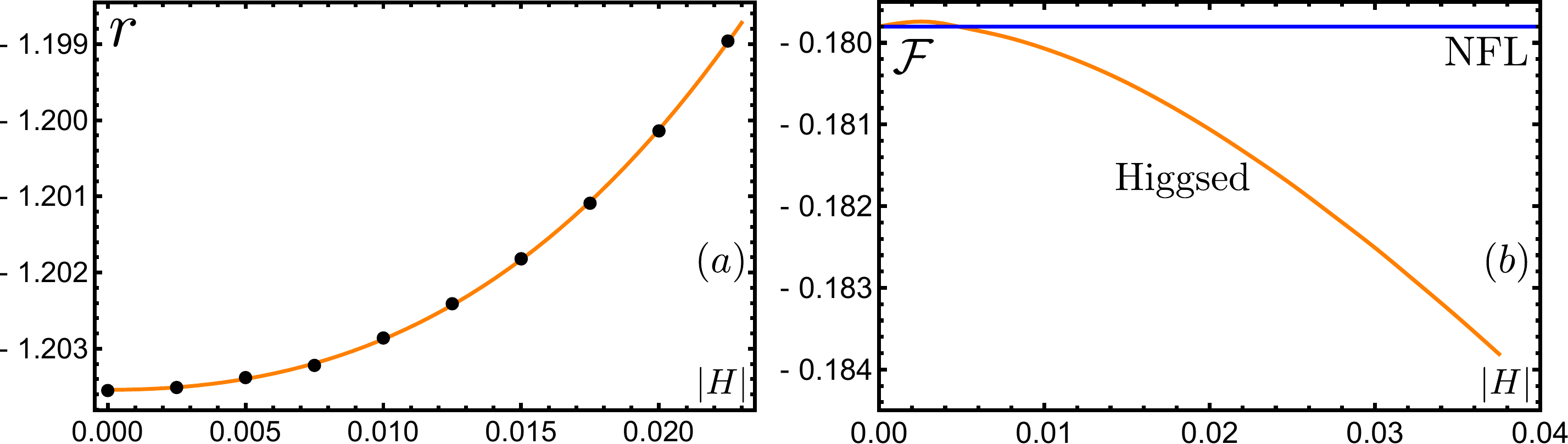} 
\end{center}
\caption{(a) Plot of $r$ as a function of $|H|$ in the higgsed phase, obtained from numerical solution of (\ref{Dysonhiggs}) in the $T\rightarrow0$ limit. The orange line fits the numerical data with $r=r_c+h_2|H|^2+h_3|H|^3$, so $r-r_c\sim |H|^2$ as $|H|\rightarrow0$. The values of parameters used are $t=t_H=g^2=g_H=1$, $2M=N$ and $\mu=0$. (b) Plot of the free energies per fermionic degree of freedom of the $|H|\neq0$ solution (orange) and $H=0$ solution (blue) of (\ref{Dysonhiggs}). The weak first-order behavior at very small $|H|$ is due to a small finite $T=10^{-5}$ in the numerics, and disappears as $T\rightarrow0$. The values of other parameters used are the same as in (a).}
\label{Higgstn}
\end{figure}

Fig.~\ref{Higgstn} shows the onset of the Higgs condensate, with $r-r_c\sim |H|^2$ as $|H|\rightarrow0$, indicating a continuous transition with exponent $\nu=1/2$ as $T\rightarrow0$. Also shown is the comparison of free energies of the $|H|\neq0$ solution and the $H=0$ solution of (\ref{Dysonhiggs}) for values of $r$ that allow for the higgsed phase; this shows that the $|H|\neq0$ saddle point is indeed energetically favorable as $T\rightarrow0$.

\bibliography{EandM}

\end{document}